\documentclass[preprint,aps, amsmath,amssymb]{revtex4-1}
\newcommand{\be}{\begin{equation}}
\newcommand{\ee}{\end{equation}}
\newcommand{\bea}{\begin{eqnarray}}
\newcommand{\eea}{\end{eqnarray}}
\newcommand{\ba}{\begin{array}}
\newcommand{\ea}{\end{array}}

\usepackage{graphicx}
\usepackage{epstopdf}
\usepackage{amsmath}
\usepackage{slashed}
\usepackage{mathtools}
\usepackage{slashed}
\begin{document}

\title{Spin effects in the pion holographic light-front wavefunction}

\author{Mohammad Ahmady}
\email{mahmady@mta.ca}
\affiliation{\small Department of Physics, Mount Allison University, \mbox{Sackville, New Brunswick, Canada, E4L 1E6}}
\author{Farrukh Chishtie}
\email{fchishti@uwo.ca}
\affiliation{Theoretical Research Institute, Pakistan Academy of Sciences (TRIPAS),
Islamabad 44000, Pakistan. }

\author{Ruben Sandapen}
\email{ruben.sandapen@acadiau.ca}
\affiliation{\small Department of Physics, Acadia University, Wolfville, Nova-Scotia, Canada, B4P 2R6}
\affiliation{\small Department of Physics, Mount Allison University, \mbox{Sackville, New Brunswick, Canada, E4L 1E6}}

\begin{abstract} 
We account for dynamical spin effects in the holographic light-front wavefunction of the pion in order to predict the mean charge radius, $\sqrt{\langle r^2_{\pi} \rangle}$, the decay constant, $f_{\pi}$, the spacelike electromagnetic form factor, $F_{\pi}(Q^2)$, the twist-$2$ pion Distribution Amplitude and the photon-to-pion transition form factor $F_{\gamma \pi}(Q^2)$. Using  a universal fundamental AdS/QCD scale, $\kappa= 523$ MeV, and a constituent quark mass of $330$ MeV, we find a remarkable improvement in  describing all observables.
\end{abstract}

\maketitle

\section{Introduction}
\label{Sec:Introduction}
Hadronic light-front wavefunctions (LFWFs) provide the underlying link between the fundamental degrees of freedom of QCD, i.e. quarks and gluons, and their asymptotic hadronic states. LFWFs thus encode both the physics of confinement and chiral symmetry breaking, which are fundamental, intimately related \cite{Horn:2016rip} and yet not fully understood,  emergent properties of QCD. In phenomenology, LFWFs are extremely important since all hadronic properties can, in principle, be derived from them. For example, in the exclusive decays of the $B$ meson to light mesons, which are under intense investigation at the LHCb experiment, the theoretical non-perturbative inputs, i.e. the meson decay constants, Distribution Amplitudes and transition form factors can all be computed if the LFWFs of the mesons are known. These non-perturbative inputs are in fact the major source of theoretical uncertainties in current Standard Model predictions \cite{Ali:2016gah}.

In principle, LFWFs are obtained by solving the LF Heisenberg equation for QCD: \cite{Brodsky:2014yha}
\begin{equation}
	H_{\text{QCD}}^{\text{LF}} |\Psi(P)\rangle = M^2 |\Psi(P) \rangle
\label{LFQCD} 
\end{equation} 
where $H_{\text{QCD}}^{\text{LF}}=P^+P^- -P_{\perp}^2$ is the LF QCD Hamiltonian and $M$ is the hadron mass. At equal light-front time $(x^+=0)$ and in the light-front gauge $A^+=0$, the hadron state $|\Psi(P)\rangle$ admits a Fock expansion, i.e.  
\begin{equation}
	|\Psi(P^+, \mathbf{P_{\perp}}, S_z) \rangle = \sum_{n,h_i} \int [\mathrm{d} x_i]  [\mathrm{d}^2 \mathbf{k}_{\perp i}] \frac{1}{\sqrt{x_i}}\Psi_{n}(x_i,\mathbf{k}_{\perp i},h_i) |n: x_i P^+, x_i \mathbf{P_{\perp}} + \mathbf{k}_{\perp i}, h_i \rangle
\label{Fock-expansion}	
\end{equation} 
where $\Psi_{n}(x_i, \mathbf{k}_{\perp i},h_i)$ is the LFWF of the  Fock state with $n$ constituents and the integration measures are given by
\begin{equation}
	 [\mathrm{d} x_i] \equiv \prod_{i}^{n} \mathrm{d} x_i \delta (1-\sum_{j=1}^{n} x_j) \hspace{1cm} [\mathrm{d}^2 \mathbf{k}_{i}]\equiv \prod_{i=1}^{n} \frac{\mathrm{d}^2 \mathbf{k}_i}{2(2\pi)^3} 16 \pi^3 \delta^2(\sum_{j=1}^n \mathbf{k}_i) \;.
	 \label{Int-measures}	 
\end{equation}

LFWFs are quantum mechanical probability amplitudes that depends on the momenta fraction  $x_i=k_i^+/P^+$, the transverse momenta  $\mathbf{k}_{\perp i}$, and the helicities $h_i$ of the constituents. In practice, it is difficult, if not impossible, to solve Eq. \eqref{LFQCD} since it contains an infinite number of strongly coupled integral equations.  Various approximation schemes involve truncating the Fock expansion, using discretized light-front quantization  or solving the equations in a lower number of spatial dimensions. For a review of light-front quantum field theories, we refer to \cite{Brodsky:1997de}.

A remarkable breakthrough during the last decade is the discovery by Brodsky and de T\'eramond \cite{deTeramond:2008ht,Brodsky:2006uqa,deTeramond:2005su,Brodsky:2007hb} of a higher dimensional gravity dual to a semiclassical approximation of light-front QCD. The result is a relativistic Schr\"odinger-like wave equation for mesons that can be solved analytically to predict meson spectroscopy and LFWFs in terms of a single mass scale $\kappa$. The approach can also include baryons \cite{deTeramond:2005su} and more recently has been extended to a unified framework for baryons and mesons considered as conformal superpartners \cite{Brodsky:2016rvj}. This gauge/gravity duality is referred to as light-front holography (LFH) and is reviewed in Ref. \cite{Brodsky:2014yha}.

In the semiclassical approximation, quark masses and quantum loops are neglected, the LFWFs depend on the invariant mass $\mathcal{M}^2=(\sum_{i}^n k_i)^2$ of the constituents rather than on their individual momenta $k_i$. For the valence ($n=2$ for mesons) Fock state, the invariant mass of the $q\bar{q}$ pair is $\mathcal{M}_{q\bar{q}}^2=k_{\perp}^2/x(1-x)$ and the latter is the Fourier conjugate to the impact variable $\zeta^2=x(1-x) b^2$ where $b$ is the transverse separation the quark and antiquark. The valence meson LFWF can then be written in a factorized form: 
\begin{equation}
	\Psi(\zeta, x, \phi)= e^{iL\phi} \mathcal{X}(x) \frac{\phi (\zeta)}{\sqrt{2 \pi \zeta}} 
\label{mesonwf}
\end{equation}
where the helicity indices have been suppressed \cite{Brodsky:2014yha}. We note that this suppression of the helicity indices is legitimate if either the constituents are assumed to be spinless or if the helicity dependence decouples from the dynamics. It can then be shown that Eq. \eqref{LFQCD} reduces to a $1$-dimensional Schr\"odinger-like wave equation for the transverse mode of LFWF of the valence ($n=2$ for mesons) state, namely: 
\begin{equation}
 	\left(-\frac{d^2}{d\zeta^2} - \frac{1-4L^2}{4\zeta^2} + U(\zeta) \right) \phi(\zeta) = M^2 \phi(\zeta) 
 	\label{hSE}
 \end{equation}
where all the interaction terms and the effects of higher Fock states on the valence state are hidden in the confinement potential $U(\zeta)$. The latter remains to be specified and, at present, this cannot be done from first principles in QCD. 

However, Brodsky and de T\'eramond found that Eq. \eqref{hSE} maps onto the wave equation for the propagation of spin-$J$ string modes in the higher dimensional anti-de Sitter space, $\text{AdS}_5$, if the impact light-front variable $\zeta$ is identified with $z_5$, the fifth dimension of AdS space and the light-front orbital angular momentum $L^2$ is mapped onto $(mR)^2-(2-J)^2$  where $R$ and $m$ are the AdS radius and mass respectively. For this reason, we refer to Eq. \eqref{hSE} as the holographic LF Schr\"odinger equation. In this AdS/QCD duality, the confining potential in physical spacetime is driven by the deformation of the pure $\text{AdS}_5$ geometry.  Specifically, the potential is given by
 \begin{equation}
 	U(z_5, J)= \frac{1}{2} \varphi^{\prime\prime}(z_5) + \frac{1}{4} \varphi^{\prime}(z_5)^2 + \left(\frac{2J-3}{4 z_5} \right)\varphi^{\prime} (z_5) 
 \end{equation}
 where $\varphi(z_5)$ is the dilaton field which breaks  conformal invariance in AdS space. A quadratic dilaton, $\varphi(z_5)=\kappa^2 z_5^2$, profile results in a light-front harmonic oscillator potential in physical spacetime:
 \begin{equation}
 	U(\zeta,J)= \kappa^4 \zeta^2 + \kappa^2 (J-1) 
 	\label{harmonic-LF}
 \end{equation}
 since $z_5$ maps onto the LF impact variable $\zeta$. Remarkably, Brodsky, Dosch and de T\'eramond have shown  that the quadratic form of the AdS/QCD potential is unique \cite{Brodsky:2013ar}. In fact, starting with a more general dilaton profile $\varphi \propto z_5^s$ and requiring the pion to be massless, uniquely fixes $s=2$ \cite{Brodsky:2013npa}. More formally, applying the mechanism of  de Alfaro, Furbini and Furlan \cite{deAlfaro:1976vlx} (which allows the emergence of a mass scale in the Hamiltonian of a conformal $1$-dimensional QFT while retaining the conformal invariance of the underlying action) to semiclassical LF QCD uniquely fixes the quadratic form of the  AdS/QCD potential \cite{Brodsky:2013ar}.
 
With the confining potential specified, one can solve the holographic Schr\"odinger equation to obtain the meson mass spectrum,
 \begin{equation}
 	M^2= 4\kappa^2 \left(n+L +\frac{S}{2}\right)\;
 	\label{mass-Regge}
 \end{equation}
 which, as expected, predicts a massless pion. The corresponding normalized  eigenfunctions are given by
 \begin{equation}
 	\phi_{nL}(\zeta)= \kappa^{1+L} \sqrt{\frac{2 n !}{(n+L)!}} \zeta^{1/2+L} \exp{\left(\frac{-\kappa^2 \zeta^2}{2}\right)} L_n^L(x^2 \zeta^2) \;.
 \label{phi-zeta}
 \end{equation}
 To completely specify the holographic meson wavefunction,  we need the analytic form of the longitudinal mode $\mathcal{X}(x)$.  This is obtained by matching the expressions for the pion EM or gravitational form factor in physical spacetime and in AdS space. Either matching consistently results in $\mathcal{X}(x)=\sqrt{x(1-x)}$ \cite{deTeramond:2008ht,Brodsky:2008pf}. The meson holographic LFWFs for massless quarks can thus be written in closed form:
\begin{equation}
\Psi_{nL}(\zeta, x, \phi)= e^{iL\phi} \sqrt{x(1-x)} (2\pi)^{-1/2}\kappa^{1+L} \sqrt{\frac{2 n !}{(n+L)!}} \zeta^{L} \exp{\left(\frac{-\kappa^2 \zeta^2}{2}\right)} L_n^L(x^2 \zeta^2)
\end{equation}
with the corresponding meson masses lying on linear Regge trajectories as given by Eq. \eqref{mass-Regge}. The reasons why a solution to a quantum field theory could reduce to a solution of a simple, one-dimensional differential equation are explored in Ref. \cite{Glazek:2013jba}. 

For phenomenological applications, it is necessary to restore both the quark mass and helicity dependence of the holographic LFWF. In fact, it has recently been shown in Ref. \cite{Swarnkar:2015osa} that non-zero light quark masses drastically improves the description of data on the photon-to-pion, photon-to-$\eta$ and photon-to-$\eta^{\prime}$ transition form factors. On the other hand, for the pion and kaon EM form factors, the description of data actually worsens unless $3$ data points at large $Q^2$ are excluded for the kaon form factor \cite{Swarnkar:2015osa}. Accounting for non-zero light quark masses means going beyond the semiclassical approximation, and this is usually done following the prescription of Brodsky and de T\'eramond \cite{Brodsky:2008pg}. For the ground state pion, this leads to 
\be  
\Psi^{\pi} (x,\zeta^2) = \mathcal{N} \sqrt{x (1-x)}  \exp{ \left[ -{ \kappa^2 \zeta^2  \over 2} \right] }
\exp{ \left[ -{m_f^2 \over 2 \kappa^2 x(1-x) } \right]}
\label{hwf}
\ee
where $\mathcal{N}$ is a normalization constant fixed by requiring that
 \begin{equation}
 	\int \mathrm{d}^2 \mathbf{b} \mathrm{d} x |\Psi^{\pi}(x,\zeta^2)|^2 = P_{q\bar{q}} 
 	\label{norm}
 \end{equation}
 where $P_{q\bar{q}}$ is the probability that the meson consists of the leading quark-antiquark Fock state.  

Note that Eq. \eqref{mass-Regge} tells us that the AdS/QCD scale $\kappa$ can be chosen to fit the experimentally measured Regge slopes. Ref. \cite{Brodsky:2014yha} reports $\kappa=590$ MeV for pseudoscalar mesons and $\kappa=540$ MeV for vector mesons.  A recent fit to the  Regge slopes of mesons and baryons, treated as conformal superpartners, yields  $\kappa=523$ MeV \cite{Brodsky:2016rvj}. On the other hand, the AdS/QCD scale $\kappa$ can be connected to the scheme-dependent pQCD renormalization scale $\Lambda_{\text{QCD}}$ by matching the running strong coupling in the non-perturbative (described by light-front holography) and the perturbative regimes \cite{Brodsky:2010ur}. With $\kappa=523$ MeV and the $\beta$-function of the QCD running coupling at $5$-loops,  Brodsky, Deur and de T\'eramond recently predicted the QCD renormalization scale, $\Lambda^{{\overline{MS}}}_{\text{QCD}}$, in excellent agreement with the world average value \cite{Deur:2016opc}. Furthermore, light-front holographic wavefunctions have also been used to predict diffractive vector meson production\cite{Forshaw:2012im,Ahmady:2016ujw}. A fit to the HERA data on diffractive $\rho$ electroproduction, with $m_{u/d}=140$ MeV, gives $\kappa=560$ MeV\cite{Forshaw:2012im} and using $\kappa=550$ MeV (with $m_{u/d}[m_s]=46[140]$ MeV) leads to a good simultaneous description of the HERA data on diffractive $\rho$ and $\phi$ electroproduction \cite{Ahmady:2016ujw}. These findings hint towards the emergence of a  universal fundamental AdS/QCD scale $\kappa \sim 550$ MeV. In the most recent application of LFH to predict nucleon EM form factors \cite{Sufian:2016hwn}, it is pointed out that this universality holds up to $10\%$ accuracy. In this paper, we shall use the value of $\kappa=523$ MeV which fits the meson/baryon Regge slopes and accurately predicts $\Lambda^{{\overline{MS}}}_{\text{QCD}}$ \cite{Brodsky:2016rvj, Deur:2016opc}.

In earlier applications of LFH with massless quarks, much lower values of $\kappa$ were required to fit the pion data: $\kappa=375$ MeV in Ref. \cite{Brodsky:2007hb} in order to fit the pion EM form factor data and $\kappa=432$ MeV (with $P_{q\bar{q}}=0.5$) to fit the photon-to-pion transition form factor data simultaneously at large $Q^2$ and $Q^2=0$ (the latter is fixed by the $\pi^0 \to \gamma \gamma$ decay width) \cite{Brodsky:2011xx}.  Note that in Ref. \cite{Brodsky:2007hb}, the EM form factor is computed, both in the spacelike and timelike regions, as a convolution of normalizable hadronic modes with a non-normalizable EM current which propagates in the modified infrared region of AdS space and generates the non-perturbative pole structure of the EM form factor in the timelike region. Alternatively, the spacelike EM form factor can be computed using the Drell-Yan-West formula \cite{Drell:1969km,West:1970av} in physical spacetime with the holographic pion LFWF. The latter approach is taken in Refs. \cite{Vega:2008te,Swarnkar:2015osa,Vega:2009zb,Branz:2010ub}. In  Ref. \cite{Vega:2009zb}, a higher value of $\kappa=787$ MeV is used with $m_{u/d}=330$ MeV and the authors predict $P_{q\bar{q}}=0.279$, implying an important contribution of higher Fock states in the pion. In Ref. \cite{Branz:2010ub}, a universal AdS/QCD scale $\kappa=550$ MeV is used for all mesons, together with a constituent quark mass $m_{u/d}=420$ MeV, but $P_{q\bar{q}}=0.6$ is fixed for the pion only: for the kaon, $P_{q\bar{q}}=0.8$ and for all other mesons, $P_{q\bar{q}}=1$. More recently, in Ref. \cite{Swarnkar:2015osa}, with $m_{u/d}=330$ MeV, the authors use a universal $\kappa=550$ MeV for all mesons but fix the wavefunction normalization for the pion so as to fit the decay constant. Consequently, this implies that $P_{q\bar{q}}=0.61$ only for the pion.

All these previous studies seem to indicate that a special treatment is required at least for the pion either by using a distinct AdS/QCD scale $\kappa$ or/and relaxing the normalization condition on the holographic wavefunction, i.e. invoking higher Fock states contributions. This may well be reasonable since the pion is indeed unnaturally light and does not lie on a Regge trajectory, as pointed out in Ref. \cite{Vega:2008te}. However, we note that in the previous studies \cite{Swarnkar:2015osa,Vega:2009zb,Branz:2010ub} where the pion observables are predicted using the holographic wavefunction, given by Eq. \eqref{hwf}, the helicity dependence of the latter is always assumed to decouple from the dynamics, i.e. the helicity wavefunction is taken to be momentum-independent. This is actually consistent with the semi-classical approximation within which the AdS/QCD correspondence is exact. Consequently, Ref. \cite{Branz:2010ub} derives a single formula to predict simultaneously the vector and pseudoscalar meson decay constants, so that using a universal scale $\kappa$ and $P_{q\bar{q}}=1$ for all mesons inevitably leads to degenerate decay constants in conflict with experiment.

In this paper, we show that it is possible to achieve a better description of the pion observables by using a universal AdS/QCD scale $\kappa$ and without the need to invoke higher Fock state contributions. We do so by taking into account dynamical spin effects in the holographic pion wavefunction, i.e. we use a momentum-dependent helicity wavefunction.   This approach goes beyond the semiclassical approximation, just like the inclusion of light quark masses in the holographic wavefunction. However, it does support the idea of the emergence of a universal, fundamental AdS/QCD scale. A similar approach was taken previously for the $\rho$ meson, leading to impressive agreement to the HERA data on diffractive $\rho$ electroproduction \cite{Forshaw:2012im}.

\section{Dynamical spin effects}
\label{Sec:Dynamical spin}
To restore the helicity dependence of the holographic wavefunction, we assume that
\begin{equation}
\Psi(x,\mathbf{k}) \to	\Psi_{h\bar{h}}(x, \mathbf{k}) = S_{h\bar{h}}(x, \mathbf{k})  \Psi(x,\mathbf{k})
\label{Spin-space}	
\end{equation}
where $S_{h\bar{h}}(x,\mathbf{k})$ corresponds to the helicity wavefunction for a point-like meson-$q\bar{q}$ coupling. For vector mesons,  the helicity wavefunction is therefore similar to that of the point-like photon-$q\bar{q}$ coupling, i.e. 
\begin{equation}
	S_{h\bar{h}}^{V}(x, \mathbf{k})=\frac{\bar{v}_{\bar{h}}((1-x)P^+, -\mathbf{k})}{\sqrt{(1-x)}} [\gamma \cdot \epsilon_V] \frac{u_{h}(xP^+, \mathbf{k})}{\sqrt{x}}
	\label{spinwf-vector}
	\end{equation}
 where $\epsilon_V^{\mu}$ is the polarization vector of the vector meson. Indeed, substituting $\epsilon_V^{\mu}$ by the photon polarization vector and multiplying Eq. \eqref{spinwf-vector} by the light-front energy denominator \cite{Lepage:1980fj} yields the well-known photon light-front wavefunctions \cite{Lepage:1980fj,Dosch:1996ss,Kulzinger:1998hw,Nemchik:1996cw,Forshaw:2003ki}. This assumption for the helicity structure of the vector meson is very common when computing diffractive vector meson production in the dipole model \cite{Nemchik:1996cw,Forshaw:2003ki,Watt:2007nr,Forshaw:2006np,Forshaw:2010py,Forshaw:2011yj,Armesto:2014sma,Rezaeian:2013tka,Goncalves:2015poa,Rezaeian:2012ji} and, as we mentioned earlier, was used in Ref. \cite{Forshaw:2012im} with the holographic wavefunction for the $\rho$ meson. 
 
For the pseudoscalar pion, we replace  $\gamma \cdot \epsilon_V$ in Eq. \eqref{spinwf-vector} by $(\mbox{scalar function}) \times \gamma^5$ where the most general, dimensionally homogeneous, scalar function that can be constructed using the pion's momentum is $A (P \cdot \gamma)  + B \sqrt{P \cdot P}$ with $A$ and $B$ being arbitrary constants. Hence 
 \begin{equation}
	S^{\pi}_{h\bar{h}}(x, \mathbf{k}) =\frac{\bar{v}_{\bar{h}}((1-x)P^+,-\mathbf{k})}{\sqrt{1-x}} \left[(A \slashed{P}  + B M_{\pi}) \gamma^5 \right] \frac{u_{h}(xP^+,\mathbf{k})}{\sqrt{x}} \;.
	\label{spinwf}
\end{equation}
We note that References \cite{Heinzl:2000ht,Heinzl:2000es} take $A=B=1$, quoting \cite{Dziembowski:1987zp,Jaus:1989au,Ji:1990rd,Ji:1992yf}. References \cite{Choi:2007yu,Trawinski:2016ygn} take $A=0$ and the recent paper \cite{Chang:2016ouf} considers $A=B=1$ but retains only the $\gamma^+ \gamma^5$ term in the scalar product $\slashed{P}\gamma^5$. This implies a momentum-independent (non-dynamical) helicity wavefunction if $B=0$ and that dynamical spin effects are only allowed if $B \ne 0$. After evaluating the right-hand-side using the light-front spinors given in Ref. \cite{Lepage:1980fj}, we obtain
\begin{equation}
S_{h\bar{h}}^{\pi}(x,\mathbf{k}) = \left \{ A M_{\pi}^2 + B \left(\frac{m_f M_{\pi}}{x(1-x)} \right)\right \}(2h) \delta_{-h \bar{h}} + B \left( \frac{M_{\pi} k e^{i(2h) \theta_k}}{x(1-x)}\right) \delta_{h \bar{h}}	
\label{RSpin}
\end{equation}
with $\mathbf{k}=k e^{i\theta_{k}}$. As mentioned above, if we take $B=0$, the helicity wavefunction becomes momentum-independent: 
	\begin{equation}
	S^{\pi}_{h\bar{h}}(x, \mathbf{k}) \to	S^{\pi}_{h\bar{h}} = \frac{1}{\sqrt{2}}  (2h) \delta_{-h \bar{h}}
	\label{NRspin}
		\end{equation}
		normalized such that $\sum_{h\bar{h}} |S^{\pi}_{h,\bar{h}}|^2=1$. Such a helicity wavefunction is assumed for the  meson (both pseudoscalar and vector) holographic wavefunction in Refs. \cite{Swarnkar:2015osa,Vega:2009zb} and we shall refer to it as the non-dynamical (i.e. momentum-independent) helicity wavefunction, consistent with the semi-classical approximation of light-front holography. Our spin-improved helicity wavefunction allows for an additional momentum-dependent contribution in the opposite-helicities part of the wavefunction as well as configurations in which the quark and antiquark have same helicities  \cite{Leutwyler:1973mu,Leutwyler:1973tn}.  Note that the same-helicities terms are eigenfunctions of the LF orbital angular momentum operator given by \cite{Brodsky:2014yha}
\begin{equation}
	L_z = -i \left(k_y \partial_{k_{x}} - k_x \partial_{k_{y}} \right) = i \partial_{\theta_{k}}	\end{equation}
with eigenvalues $L_z=-2h$. In other words, for this same-helicities component of our pion wavefunction, the orbital angular momentum $L_z=-S_z$ where $S_z=h +\bar{h}=2 h$ so that $J_z=L_z+ S_z=0$ as required for the pion. Note that when we allow for dynamical spin effects, we are going beyond the semi-classical approximation, and Eq. \eqref{mass-Regge} needs to modified  due to a spin-orbit interaction term (not specified in this paper) in the light-front Sch\"odinger equation. It is also useful to check that our spin-improved wavefunction transforms correctly under the LF parity operator, $\mathcal{P}_{\perp}$, which flips the signs of all helicities and that of the $x$ (or $y$) component of the transverse momentum: \cite{Brodsky:2006ez}
\begin{equation}
	\Psi^{\pi}_{h \bar{h}} (x, \mathbf{k}) \xrightarrow{\mathcal{P} _{\perp}} \Psi^{\pi}_{-h,-\bar{h}} (x, \tilde{\mathbf{k}}) 	
	\label{Parity}
	\end{equation}
where $\tilde{\mathbf{k}}=-k_x + i k_y=k e^{i(\pi-\theta_k)}$.  Since $e^{-i2h(\pi-\theta_k)}=-e^{i2h\theta_k}$, it is explicit from Eq. \eqref{RSpin}, that our spin-improved wavefunction is parity-odd, i.e.
\begin{equation}
	\mathcal{P}_{\perp} \Psi^{\pi}_{h,\bar{h}} (x, \mathbf{k}) = -\Psi^{\pi}_{h,\bar{h}} (x, \mathbf{k})
	\label{odd-parity}
	\end{equation}
as required for the pion. 

A two-dimensional Fourier transform of our spin-improved wavefunction to impact space gives 
\begin{equation}
 \Psi^{\pi}_{h\bar{h}}(x, \mathbf{b}) =	 \{ (A x(1-x) M_{\pi}^2  + B m_f M_{\pi})  (2h) \delta_{-h \bar{h}} -  B M_{\pi} i \partial_b \delta_{h \bar{h}}  \}  \frac{\Psi^{\pi} (x,\zeta^2)}{x(1-x)}  
\label{spin-improved}		
\end{equation}
which can be compared to the original holographic wavefunction, 
\begin{equation}
\Psi^{\pi[\text{o}]}_{h\bar{h}}(x,\mathbf{b})=  \frac{1}{\sqrt{2}}  h \delta_{-h \bar{h}} \Psi^{\pi}(x,\zeta^2) 
\label{original}
\end{equation}
where $\Psi^{\pi}(x,\zeta^2)$ in both of the above equations, is the holographic wavefunction given by Eq. \eqref{hwf}. We now fix the normalization constant $\mathcal{N}$ appearing in Eq. \eqref{hwf} by requiring that 
\begin{equation}
 	 \int \mathrm{d}^2 \mathbf{b} \mathrm{d} x |\Psi^{\pi} (x,\mathbf{b})|^2 = 1 
 	\label{normhbh}
 \end{equation}
 where
 \begin{equation}
 	|\Psi^{\pi}(x,\mathbf{b})|^2 \equiv \sum_{h,\bar{h}} 
 	|\Psi^{\pi}_{h\bar{h}} (x,\mathbf{b})|^2 \;.
 	\label{sum-notation}
 	\end{equation}
Note that Eq. \eqref{normhbh} reduces to the normalization condition given by Eq. \eqref{norm} (with $P_{q\bar{q}}=1$) if we substitute the original holographic wavefunction, Eq. \eqref{original}, in Eq. \eqref{normhbh}. Imposing our normalization condition, Eq. \eqref{normhbh}, implies that we assume that the pion consists only of the leading quark-antiquark Fock state. 

\begin{figure}[htbp]
\centering 
\includegraphics[width=14cm,height=14cm]{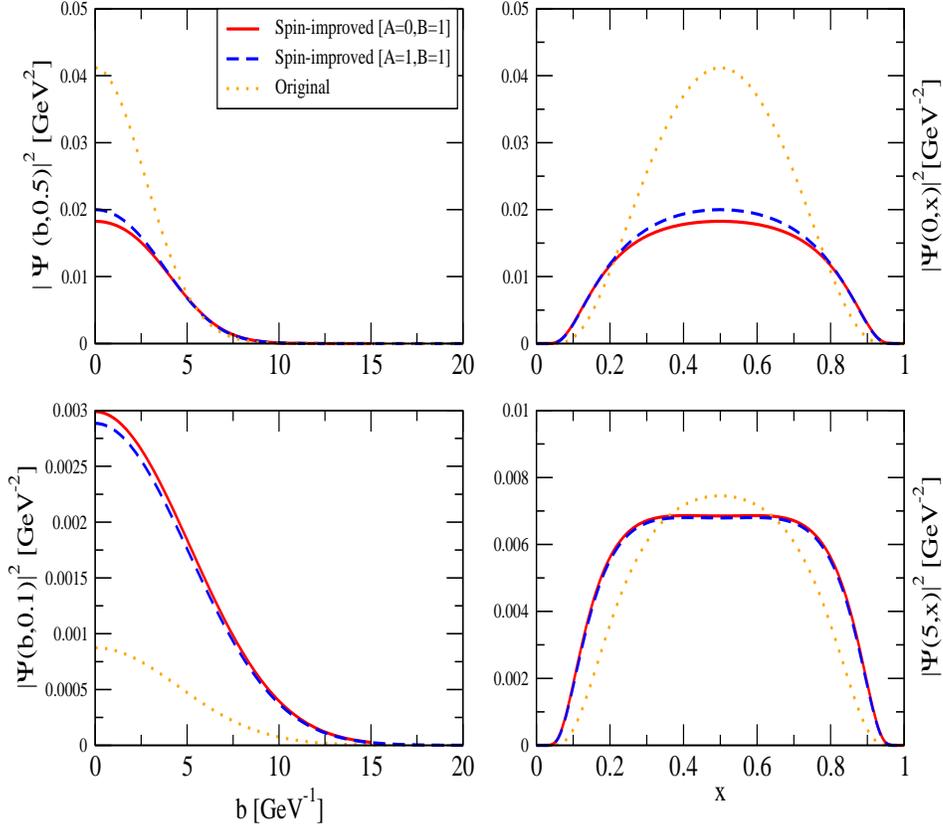}
\caption{The pion holographic LFWF squared and summed over all helicities: $|\Psi^{\pi}(x,\mathbf{b})|^2 ~[\text{GeV}^{2}]$. Dotted-orange: original. Continous-red: spin-improved ($A=0,B=1$). Dashed-blue: spin-improved ($A=1,B=1$		). Left: The $b$-dependence of the wavefunction at fixed $x=0.5$ (upper)  and $x=0.1$ (lower). Right: The $x$-dependence of the wavefunction at fixed $b=0$ (upper) and $b=5~\text{GeV}^{-1}$ (lower). All plots are generated with $\kappa=523$ MeV and $m_{u/d}=330$ MeV.}
\label{fig:wfs}
\end{figure}

\begin{figure}[htbp]
\centering 
\includegraphics[width=10cm,height=10cm]{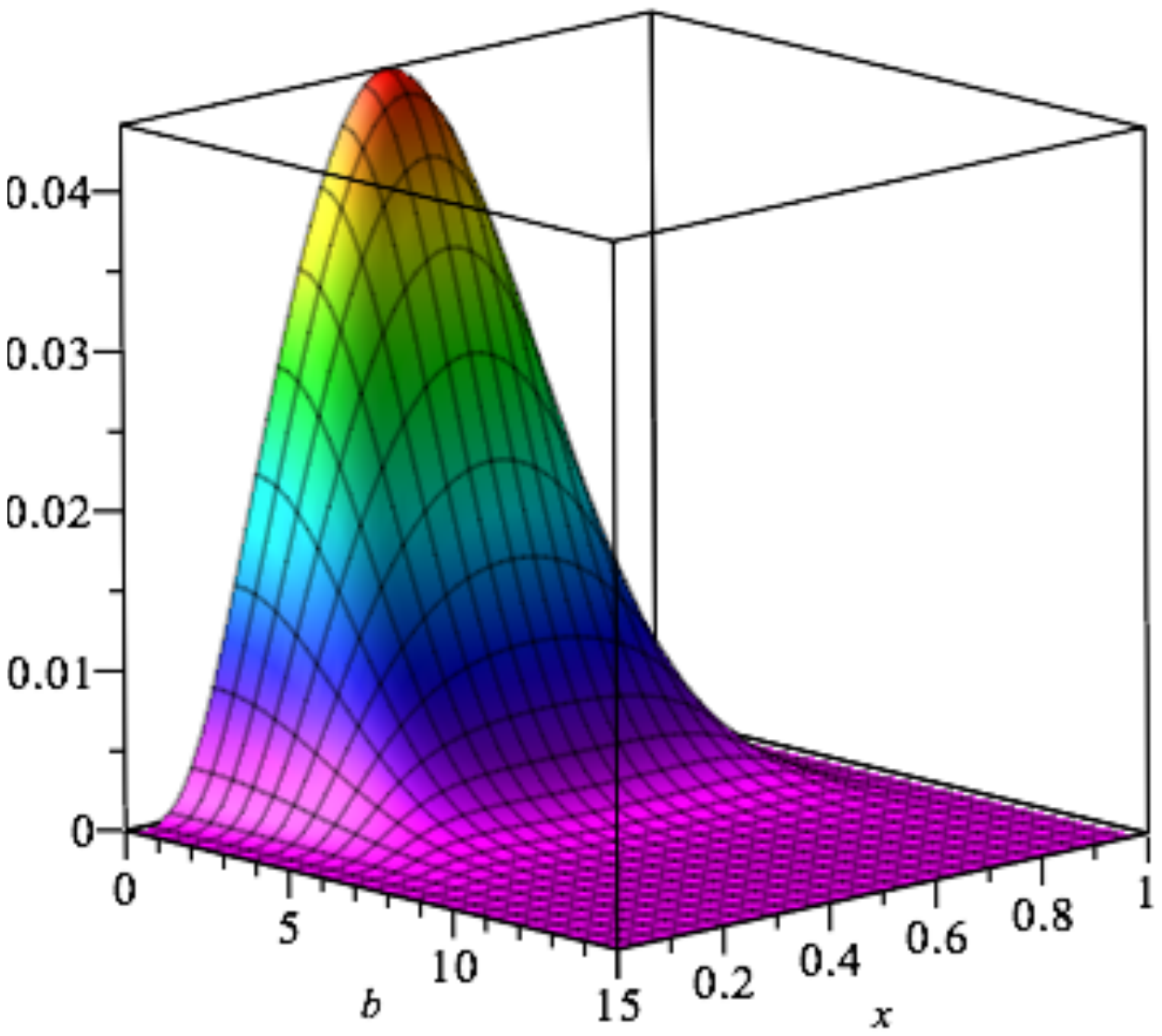}
\vspace{-4cm}
\includegraphics[width=10cm,height=10cm]{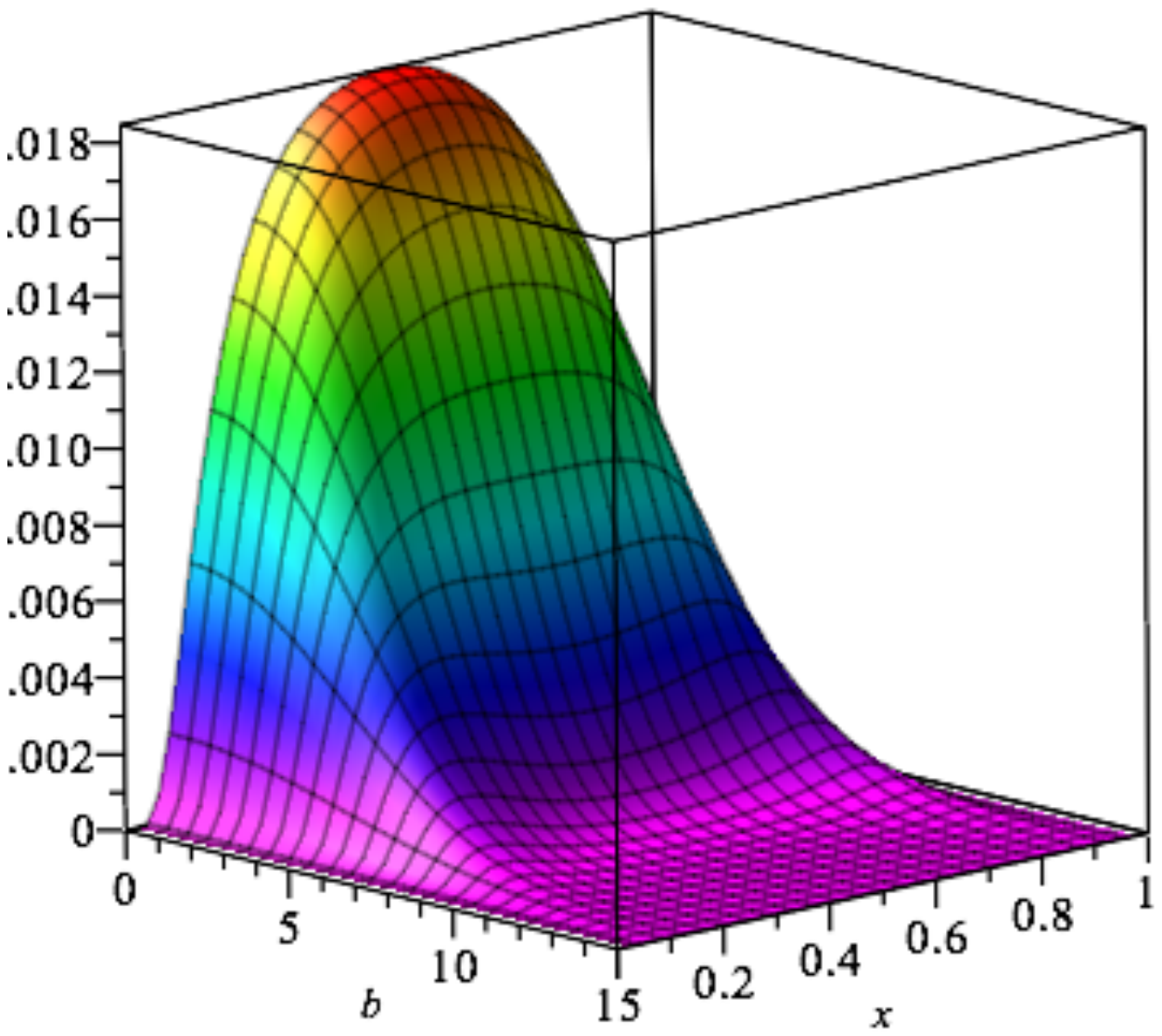}\includegraphics[width=10cm,height=10cm]{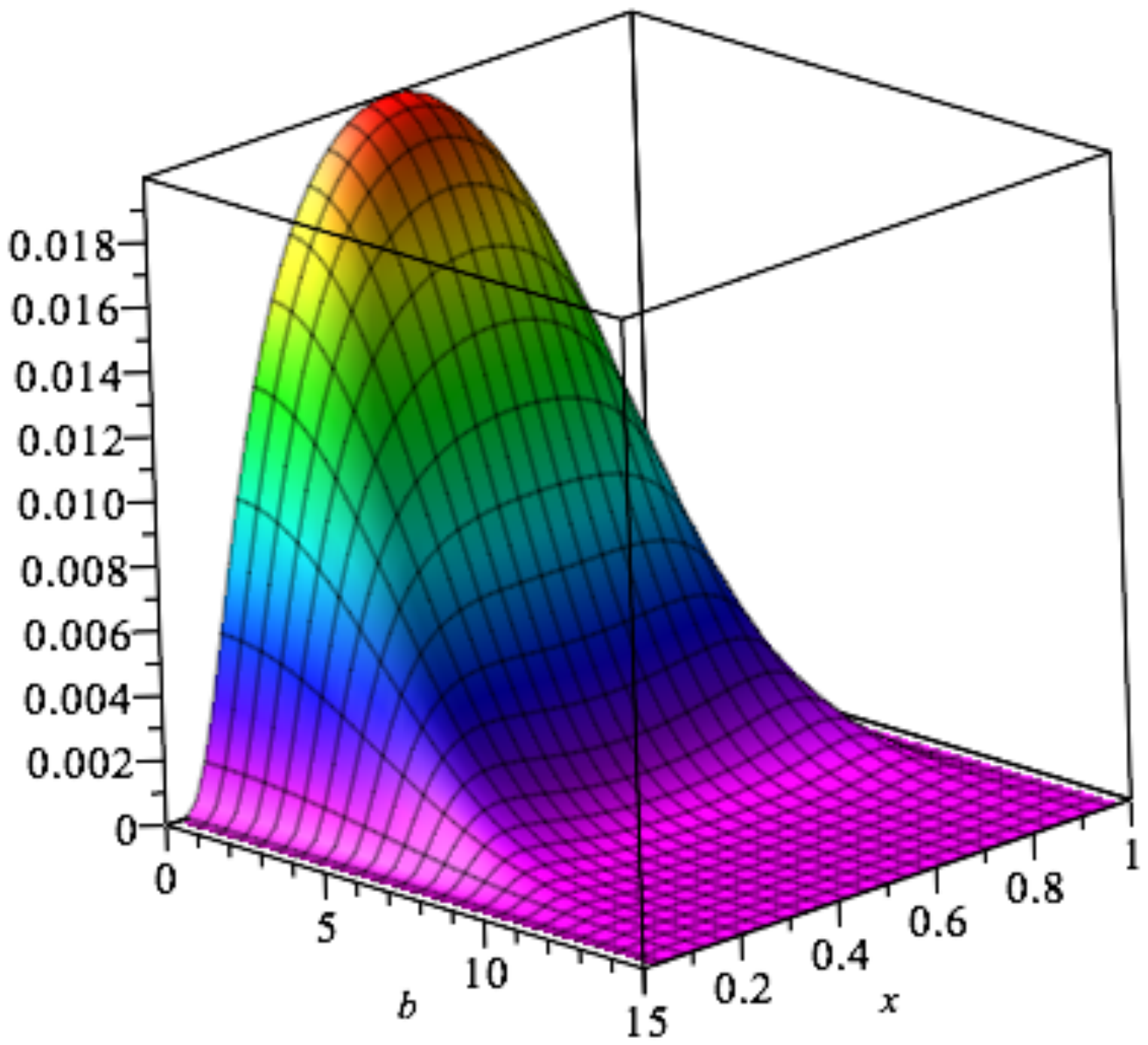}
\caption{The normalized pion holographic wavefunction squared and summed over helicities, $|\Psi^{\pi}(x,\mathbf{b})|^2 ~ [\text{GeV}^{2}]$, as a function of the transverse separation $b~[\text{GeV}^{-1}]$  and momentum fraction $x$. Upper: original. Lower left: spin-improved ($A=0,B=1$). Lower right: spin-improved ($A=1,B=1$). All plots are generated with $\kappa=523$ MeV and $m_f=330$ MeV.} 
\label{fig:3dwf}
\end{figure}

In Figure \ref{fig:wfs}, we show the dynamical spin effects in the squared and helicity-summed holographic wavefunction with a constituent quark mass, $m_{u/d}=330$ MeV. Recall that we recover the original holographic pion wavefunction by taking $B=0$ in our spin-improved wavefunction. In addition, we consider the two cases $[A=0,B=1]$ and $[A=1,B=1]$ that allow for dynamical spin effects. It can be seen that, at fixed $x=0.5$ (and $x=0.1$), the spin-improved wavefunctions are suppressed (and enhanced) respectively, compared to the original wavefunction. At fixed $b=0$ and $b=5~\mbox{GeV}^{-1}$, the spin-improved wavefunctions are broader than the original wavefunction. In Figure \ref{fig:3dwf}, we compare the $3$-dimensional plots of the spin-improved wavefunctions to the original wavefunction, which clearly show that dynamical spin effects enhance the end-point contributions in $x$. 
 
\section{Radius and decay constant}
Having specified our spin-improved holographic wavefunction, we shall now compute two observables: the pion radius, sensitive to long-distance (non-perturbative) physics and the pion decay constant, sensitive to short-distance (perturbative) physics. We shall predict both observables using the original and spin-improved holographic wavefunctions with a constituent quark mass, $m_{u/d}=330$ MeV. We expect to fit better the radius since the holographic pion wavefunction lacks the perturbative, short-distance corrections that may be required to accurately predict the decay constant.  

The root-mean-square pion radius is given by \cite{Brodsky:2007hb}:
\begin{equation}
	\sqrt{\langle r_{\pi}^2 \rangle} = \left[\frac{3}{2} \int \mathrm{d} x \mathrm{d}^2 \mathbf{b} [b (1-x)]^2 |\Psi^{\pi}(x,\mathbf{b})|^2 \right]^{1/2} 
	\label{radius}
\end{equation}
where $|\Psi^{\pi}(x,\mathbf{b})|^2$ is given by Eq. \eqref{sum-notation}. Our predictions for the pion radius are compared to the measured value in Table \ref{tab:radius}.  As can be seen, we achieve a much better agreement with the datum with the spin-improved holographic wavefunctions. It is worth noting the excellent agreement achieved with the ($A=1,B=1$) spin-improved wavefunction.

%%%%%%%%%%%%%%%%%%%%%%%%%%%%%%%%%
\begin{table}
  \centering
  \begin{tabular}{|c|c|}
    \hline
    & $\sqrt{\langle r_{\pi}^2 \rangle}$ [fm]\\
        \hline
    Original  & $0.544$\\
        \hline
   Spin-improved ($A=0,B=1$) & $0.683$ \\
    \hline
   Spin-improved ($A=1,B=1$)& $0.673$ \\
   \hline
    Experiment \cite{Agashe:2014kda}   & $0.672 \pm 0.008$\\
    \hline
    \end{tabular}
  \caption{Our predictions for the pion radius using the holographic wavefunction with $\kappa=523$ MeV and $m_{u/d}=330$ MeV. The datum is from PDG 2014 \cite{Agashe:2014kda}.}
  \label{tab:radius}
\end{table}
%%%%%%%%%%%%%%%%%%%%%%%%%%%%%%%%
 
Note that if we compute the pion radius using the original holographic wavefunction but with $\kappa=540$ MeV and $m_f=330$ MeV as in Ref. \cite{Swarnkar:2015osa}, we obtain $\sqrt{\langle r_{\pi}^2 \rangle}=0.530$ fm which is to be compared with the prediction of  Ref. \cite{Swarnkar:2015osa}: $\sqrt{\langle r_{\pi}^2 \rangle}=0.529$ fm. In Ref. \cite{Swarnkar:2015osa}, the authors obtain the pion radius from the slope of the EM pion form factor $F_{\pi}(Q^2)$ at $Q^2=0$ with the constraint that $F(0)=1$. We note that the latter constraint on the form factor is automatically satisfied if the pion wavefunction is normalized with $P_{q\bar{q}}=1$. Thus, although the authors of Ref. \cite{Swarnkar:2015osa} imply $P_{q\bar{q}}=0.62$ in order to fit the decay constant, they implicitly assume $P_{q\bar{q}}=1$  when computing the EM form factor. This is why we are able to reproduce their prediction for the pion radius even though we assume $P_{q\bar{q}}=1$. 
 
%For completeness, we also compute the transverse size, $R_{\perp}$, of the pion defined as \cite{Heinzl:2000ht}
%\begin{equation}
%	R_{\perp} = \frac{1}{\sqrt{\langle k^2 \rangle}}
%\end{equation}
%where
%\begin{equation}
%\langle k^2 \rangle = \int \mathrm{d} x \mathrm{d}^2 \mathbf{k} \mathbf{k}^2 \sum_{h\bar{h}} |\Psi_{h\bar{h}}(x,\mathbf{k})|^2 	
%\end{equation}
%is the mean value of the quark (or antiquark) transverse momentum squared. Our results are shown in Table \ref{tab:trans-radius}. It can be seen that for both holographic wavefunctions, we predict $\sqrt{\langle k^2 \rangle} \sim m_f > M_{\pi}$ which, as noted in Ref. \cite{Heinzl:2000ht}, confirms the highly relativistic nature of the pion. As expected, the spin-improved holographic wavefunction gives a more reasonable  prediction (as expected from the measured radius) for the transverse size of the pion. 

%Having selected the $[A=0,B=1]$ combination for the spin-improved holographic pion wavefunction, we shall now use it to predict the pion EM form factor, the pion decay constant and Distribution Amplitudes and finally the photon-to-pion transition form factor. 

%%%%%%%%%%%%%%%%%%%%%%%%%%%%%%%%%
%\begin{table}
%  \centering
%  \begin{tabular}{|c|c|c|}
%    \hline
%    &  $\sqrt{\langle k^2 \rangle}$ [MeV]&$R_{\perp}$[fm]\\
%    \hline
%    Original  & $250$ & $0.791$\\
%    \hline
%    Spin-improved  &$286$ & $0.690$\\
%    \hline
%    \end{tabular}
%  \caption{Our predictions for the mean transverse momentum and size with $\kappa=523$ MeV and $m_f=330$ MeV.}
%  \label{tab:trans-radius}
%\end{table}
%%%%%%%%%%%%%%%%%%%%%%%%%%%%%%%%
We now compute the pion decay constant, $f_{\pi}$, defined by \cite{Lepage:1980fj}
\begin{equation}
	\langle 0 | \bar{\Psi}_d \gamma^\mu \gamma_5 \Psi_u | \pi^+ \rangle =  f_{\pi} P^\mu 
\label{fpi-def}
\end{equation}
where we have omitted to write a conventional $\sqrt{2} i$ factor on the right-hand-side. Taking $\mu=+$ and expanding the left-hand-side of Eq. \eqref{fpi-def},  we obtain 
\begin{equation}
	\langle 0 | \bar{\Psi}_{d} \gamma^+ \gamma^5  \Psi_u | \pi^+ \rangle=\sqrt{4 \pi N_c} \sum_{h, \bar{h}} \int \frac{\mathrm{d}^2 \mathbf{k}}{16\pi^3} \mathrm{d} x \Psi_{h,\bar{h}}^{\pi}(x,\mathbf{k}) \left \{\frac{\bar{v}_{\bar{h}}}{\sqrt{1-x}} (\gamma^+ \gamma^5 )\frac{u_{h}}{\sqrt{x}}\right \} \;.
	\end{equation}
The light-front matrix element in curly brackets can readily be evaluated: 
\begin{equation}
\left \{\frac{\bar{v}_{\bar{h}}}{\sqrt{1-x}} (\gamma^+ \gamma^5 )\frac{u_{h}}{\sqrt{x}}\right \}=2P^+ (2h) \delta_{-h\bar{h}}
\label{+gamma5} \;,
\end{equation} 
which implies that only the opposite-helicities term in the holographic wavefunction contributes to the decay constant. We note, however, that the same-helicities term affects the normalization of our wavefunction and thus our prediction for the decay constant. Using our spin-improved wavefunction, Eq. \eqref{spin-improved}, we deduce that
\begin{equation}
	f_{\pi}= 2 \sqrt{\frac{N_c}{\pi}}  \left.\int \mathrm{d} x   \{A((x(1-x) M_{\pi}^2)+ B m_f M_{\pi}\} \frac{\Psi^{\pi} (x,\zeta)}{x(1-x)}\right|_{\zeta=0} \;.
	\label{fpi-spin}
\end{equation}
On the other hand, using the original holographic wavefunction, Eq. \eqref{original},  we obtain 
\begin{equation}
	f^{\text{[o]}}_{\pi}= 2  \sqrt{6} \sqrt{4\pi} \int \mathrm{d} x \frac{\mathrm{d}^2\mathbf{k}}{16 \pi^3} \Psi(x,\mathbf{k})= 2 \sqrt{2} \sqrt{\frac{N_c}{4\pi}}\left.\int \mathrm{d} x  \Psi^{\pi} (x,\zeta)\right|_{\zeta=0} 	\label{fpi-o}
\end{equation}
where we have written the momentum-space expression to point out that, up to a factor of $\sqrt{4\pi}$, it coincides with the  formula for the decay constant given in Ref. \cite{Lepage:1982gd} and widely used in the literature, as for example, in Refs. \cite{Chabysheva:2012fe,Swarnkar:2015osa,Branz:2010ub}. The $\sqrt{4\pi}$ factor mismatch is consistent with the fact that our normalization in momentum-space (Eq. \eqref{norm}) differs from the conventional light-front normalization \cite{Lepage:1980fj} by a factor of $4\pi$. 

Our predictions for the pion decay constant are shown in Table \ref{tab:fpi}. As can be seen, we achieve a much better agreement with the datum with the spin-improved holographic wavefunctions although we still somewhat overestimate the measured value. As we noted above, this could perhaps be attributed to the fact that perturbative corrections are not included in the holographic pion wavefunction.

%%%%%%%%%%%%%%%%%%%%%%%%%%%%%%%%%
\begin{table}
  \centering
  \begin{tabular}{|c|c|}
    \hline
    & $f_{\pi}$ [MeV] \\
    \hline
    Original &$161$ \\
    \hline
    Spin-improved ($A=0,B=1$) & $135$  \\
    \hline
    Spin-improved ($A=1,B=1$) & $138$  \\
    \hline
    Experiment \cite{Agashe:2014kda}  &  $130.4 \pm 0.04 \pm 0.2$ \\
    \hline
    \end{tabular}
  \caption{Our predictions for the pion decay constant  using the holographic wavefunction with $\kappa=523$ MeV and $m_{u/d}=330$ MeV. The datum is from PDG 2014 \cite{Agashe:2014kda}.}
  \label{tab:fpi}
\end{table}
%%%%%%%%%%%%%%%%%%%%%%%%%%%%%%%%

\section{EM form factor}
\label{Section:EMFF}
We now compute the pion EM form factor defined as
\begin{equation}
	\langle  \pi^+ : P^{\prime}| J_{\text{em}}^\mu (0) | \pi^+: P \rangle = 2 (P + P^{\prime})^\mu F_{\pi}(Q^2)
\end{equation}
where $P^{\prime}=P+q$, $Q^2=-q^2$ and the EM current $J_{\text{em}}^\mu(z)=\sum_f e_f \bar{\Psi} (z) \gamma^\mu \Psi(z)$ with $f=\bar{d},u$ and $e_{\bar{d},u}=1/3,2/3$. The EM form factor can be expressed in terms of the pion LFWF using the Drell-Yan-West formula \cite{Drell:1969km,West:1970av}:
\begin{equation}
	F_{\pi}(Q^2)= 2 \pi \int \mathrm{d} x \mathrm{d} b ~ b ~ J_{0}[(1-x)  b Q] ~ |\Psi^{\pi}(x,\textbf{b})|^2 
\label{DYW}
\end{equation}
where $|\Psi^{\pi}(x,\textbf{b})|^2$ is given by Eq. \eqref{sum-notation}. Note that Eq. \eqref{DYW} implies that $F_{\pi}(0)=1$ if the pion LFWF is normalized according to Eq. \eqref{normhbh} and that the slope of the EM form factor at $Q^2=0$ is related to the mean radius of the pion given by Eq. \eqref{radius} via 
\begin{equation}
	\langle r_{\pi}^2 \rangle = -\frac{6}{F_{\pi}(0)} \left . \frac{\mathrm{d} F_{\pi}}{\mathrm{d} Q^2} \right|_{Q^2=0} \;.
\end{equation}

\begin{figure}[htbp]
\centering 
\includegraphics[width=16cm,height=16cm]{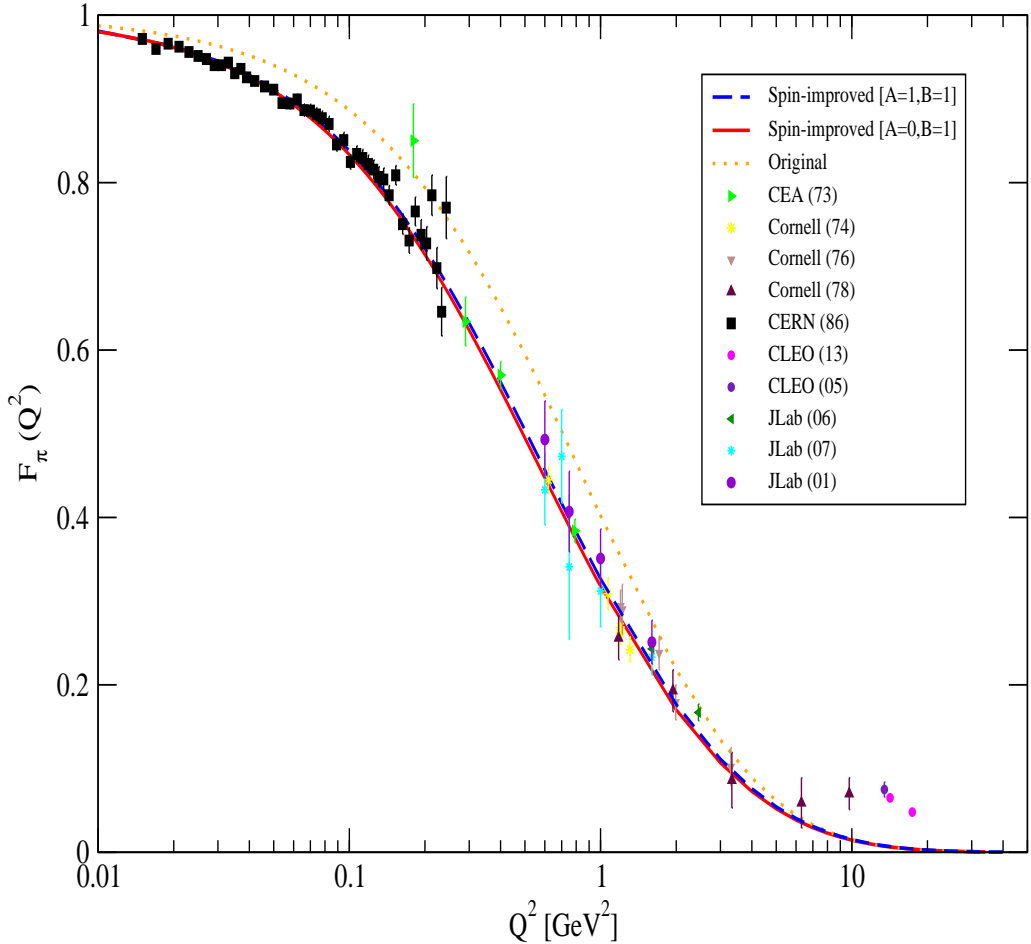}
\caption{Our predictions for the pion EM form factor. Dotted-orange: original. Continuous-red curve: spin-improved ($A=0,B=1$). Dashed-blue curve: spin-improved ($A=1,B=1$). All theory curves generated with $\kappa=523$ MeV and $m_f=330$ MeV. Data from \cite{Amendolia:1986wj,Bebek:1974,Bebek:1976,Bebek:1978,Volmer:2001,Horn:2006tm,Pedlar:2005sj,Seth:2012nn}.} 
\label{Fig:EMFF}
\end{figure}

Our predictions for the EM form factor using  the original (dotted-orange curve) and our higher twist spin-improved (continuous-red and dashed-blue curve) are compared with the data from CERN \cite{Amendolia:1986wj}, CEA \cite{Brown:1973}, Cornell \cite{Bebek:1974,Bebek:1976,Bebek:1978}, Jlab \cite{Volmer:2001,Horn:2006tm} and  CLEO \cite{Pedlar:2005sj,Seth:2012nn} in Figure \ref{Fig:EMFF}. As can be seen, the agreement with data is very much improved with the spin-improved holographic wavefunctions. In fact, we achieve excellent agreement with data from the lowest $Q^2$ datum to $Q^2 \approx 7~\text{GeV}^2$. For $Q^2 > 7~\text{GeV}^2$, our predictions with the original and the spin-improved holographic wavefunctions coincide and they both undershoot the precise CLEO data \cite{Pedlar:2005sj}. This is the short-distance regime where perturbative corrections, not taken into account in the purely non-perturbative holographic wavefunction, become important. It is worth highlighting that agreement with the precise data in the non-perturbative region, $Q^2 \le 1~\text{GeV}^2$, is excellent with our higher twist spin-improved holographic wavefunctions. 

\section{Distribution Amplitude and transition form factor}

We can also predict the twist-$2$ holographic pion DA, $\varphi_{\pi} (x,\mu)$,  defined as \cite{Radyushkin:1977gp,Lepage:1980fj}
\begin{equation}
	\langle 0 | \bar{\Psi}_d (z) \gamma^+ \gamma_5 \Psi_u (0)| \pi^+ \rangle = f_{\pi} P^+ \int \mathrm{d} x e^{i x (P \cdot z)} \varphi_{\pi} (x,\mu) 
\label{DApi}
\end{equation}
where $z^2=0$. The DA is conventionally normalized as
\begin{equation}
	\int \mathrm{d} x \varphi_{\pi}(x,\mu) = 1
	\label{DAnorm}
\end{equation}
such that taking the limit of local operators ($z\to 0$) in Eq. \eqref{DApi}, we recover the definition of the pion decay constant given by Eq. \eqref{fpi-def} (with $\mu=+$). Proceeding in the same manner as for the decay constant, we are able to show that
\begin{equation}
	f_{\pi} \varphi_{\pi}(x,\mu)=  2 \sqrt{\frac{N_c}{\pi}} \int \mathrm{d} b J_{0}(\mu b) b \{A((x(1-x) M_{\pi}^2)+ B m_f M_{\pi}\} \frac{\Psi^{\pi} (x,\zeta)}{x(1-x)} \;.
	\label{DA}
\end{equation}

%when using our spin-improved holographic wavefunction, and
%\begin{equation}
%	f^{\text{[o]}}_{\pi} \varphi^{\text{[o]}}_{\pi}(x,\mu)= \sqrt{2} \sqrt{\frac{N_c}{\pi}} \int \mathrm{d} b J_{0}(\mu b) b  \Psi (\zeta,x) \;.
%	\label{DA-o}
%\end{equation}
%when using the original holographic wavefunction. 

\begin{figure}[htbp]
\centering 
\includegraphics[width=14cm,height=14cm]{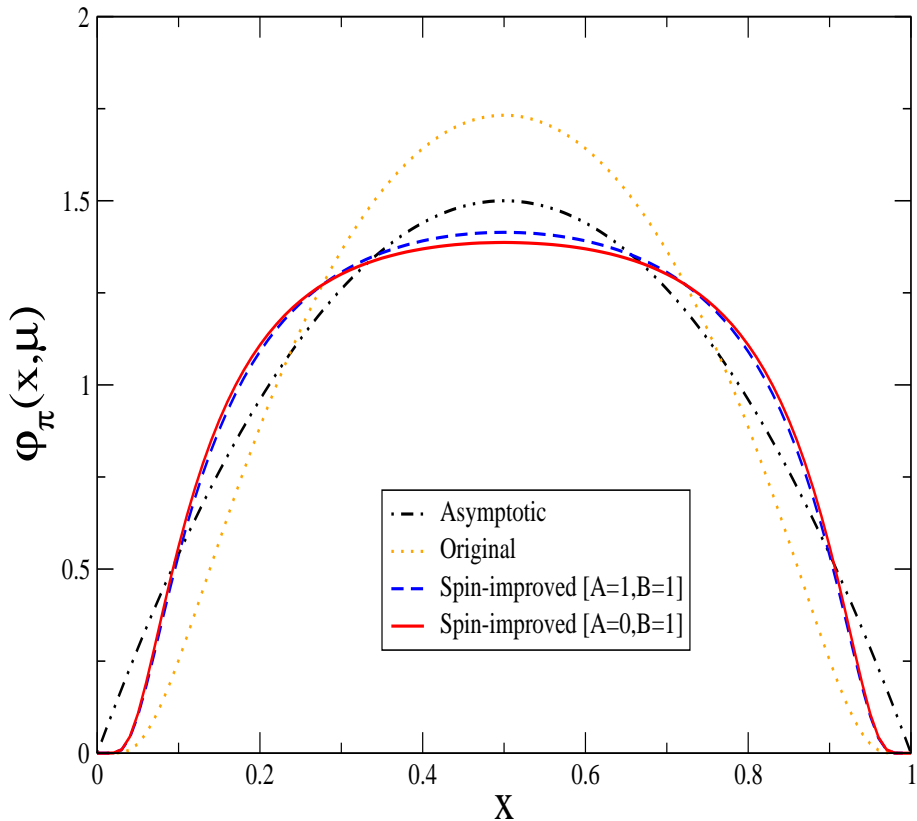}   
\caption{Comparing our spin-improved holographic DAs (continuous-red and dashed-blue curves) to the original  holographic DA (dotted-orange curve) at a scale $\mu=1$ GeV, both with $\kappa=523$ MeV and $m_f=330$ MeV. The asymptotic DA is the dotted black curve.} 
\label{fig:DA}
\end{figure}

\begin{figure}[htbp]
\centering 
\includegraphics[width=14cm,height=14cm]{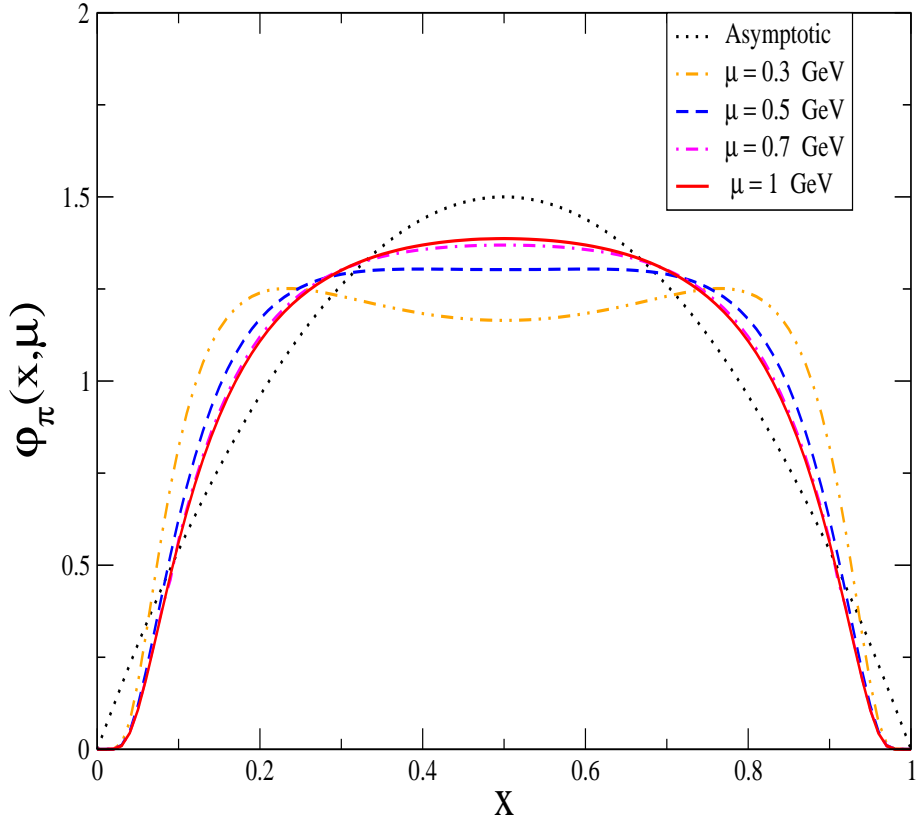}
\caption{The soft evolution of our ($(A=1,B=1)$) spin-improved holographic pion DA at a scale $\mu=0.3$ GeV(dot-dot-dashed-orange), $\mu=0.5$ GeV (dashed-blue), $\mu=0.7$ GeV (dot-dashed-magenta) and $\mu \ge 1$ GeV (continuous-red). The asymptotic DA is the dotted-black curve.} 
\label{fig:DAevo}
\end{figure}

In Figure \ref{fig:DA}, we compare our spin-improved holographic DAs  to the original holographic DA and to the asymptotic DA as predicted in pQCD: $\varphi_{\pi}(x,\infty)=6 x(1-x)$. It can be seen that our spin-improved holographic DAs (continuous-red and dashed-blue curves) are broader than both the original holographic DA (dotted-orange curve) and the asymptotic DA (dotted-black curve). All holographic DAs are both generated with $\mu=1$ GeV and we note they hardly evolve for $\mu > 1$ GeV. In other words, our holographic DAs lack the hard,  perturbative evolution given by the Efremov-Radyushkin-Brodsky-Lepage (ERBL) equations \cite{Lepage:1979zb,Efremov:1978rn,Efremov:1979qk}. In Figure \ref{fig:DAevo}, we show the soft evolution of our spin-improved $(A=0,B=1)$ holographic DA between $\mu=0.3$ GeV and $\mu=1$ GeV. Implementing the ERLB evolution, as is done in Ref. \cite{Brodsky:2011yv}, will allow our spin-improved holographic DA to evolve beyond $\mu=1$ GeV onto the asymptotic DA. 

In order to compare our holographic DAs with the predictions of standard non-perturbative methods such as lattice QCD and QCD Sum Rules, we compute the moments defined as
\begin{equation}
	\langle \xi_{n} \rangle= \int_0^1 \mathrm{d} x (2x-1)^n \varphi_{\pi}(x,\mu)
\end{equation}
and its inverse moment is given by
\begin{equation}
	\langle x^{-1} \rangle = \int_0^1 \mathrm{d}  x\frac{\varphi_{\pi}(x,\mu)}{x} \;.
\end{equation}
Our predictions for the first two non-vanishing moments $\langle \xi_2 \rangle$ and $\langle \xi_4 \rangle$ as well as the inverse moment are shown in Table \ref{tab:DAmoments}. As can be seen, with the spin-improved holographic DAs, we achieve better agreement with the predictions of lattice QCD and QCD Sum Rules. However, our predicted moments turn out to be smaller than the predictions of all non-perturbative methods cited here. Our predicted moments are also smaller than the corresponding moments of the asymptotic DA. This discrepancy could be an indication that all dynamical spin effects might not fully captured by fixing  $B=1$. 
%%%%%%%%%%%%%%%%%%%%%%%%%%%%%%%%%
\begin{table}
  \centering
  \begin{tabular}{|c|c|c|c|c|}
    \hline
  DA   & $\mu$ [GeV] &$\langle \xi_2 \rangle$ &$\langle \xi_4 \rangle$&$\langle x^{-1} \rangle$ \\
     \hline
     
     Asymptotic & $\infty$ & $0.2$ &$0.085$ & $3$ \\
    \hline 
    LFH spin-improved ($A=1,B=1$) &$\sim 1$  & $0.195$ &$0.076$&$2.74$\\
    \hline
    LFH spin-improved ($A=0,B=1$) &$\sim 1$  & $0.199$ &$0.078$&$2.76$\\
    \hline
    LFH (original) &$\sim 1$  & $0.151$ &$0.050$ &$2.50$\\
        \hline        
    LF Quark Model \cite{Choi:2007yu} & $\sim 1$ & $0.24 [0.22]$ &$0.11 [0.09]$ & \\
    \hline
    Sum Rules \cite{Ball:2004ye} & $1$ & $0.24$ & $0.11$&\\
    \hline
       Renormalon model \cite{Agaev:2005rc}& $1$ & $0.28$ & $0.13$&\\
     \hline  
     Instanton  vacuum \cite{Petrov:1998kg,Nam:2006au}  & $1$ & $0.22,0.21$ & $0.10,0.09$&\\ 
          \hline    
     Lattice \cite{Braun:2015axa,Braun:2006dg} & $2$ & $0.2361(41)(39),0.27 \pm 0.04$ & &\\
     \hline
     NLC Sum Rules \cite{Bakulev:2001pa} &$2$ &$0.248^{+0.016}_{-0.015}$  & $0.108^{+0.05}_{-0.03}$ &$3.16^{+0.09}_{-0.09}$ \\
    \hline
    Sum Rules\cite{Chernyak:1983ej}& $2$ & $0.343$ &$0.181$ &$4.25$\\
    
    \hline  
    Dyson-Schwinger[RL,DB]\cite{Chang:2013pq} & $2$ & $0.280,0.251$ &$0.151,0.128$ &$5.5,4.6$\\
    \hline 
    Platykurtic \cite{Stefanis:2014nla} & $2$ & $0.220^{+0.009}_{-0.006}$& $0.098^{+0.008}_{-0.005}$ & $3.13^{+0.14}_{-0.10}$\\
    \hline
    \end{tabular}
  \caption{Our predictions for the first two non-vanishing moments and the inverse moment of the pion holographic twist-$2$ DA with $\kappa=523$ MeV and $m_f=330$ MeV, compared to the predictions of lattice QCD by Braun et al. in 2006 \cite{Braun:2006dg} and 2015 \cite{Braun:2015axa}, QCD Sum Rules with non-local condensates by Bakulaev et al. \cite{Bakulev:2001pa}, QCD Sum Rules by Chernyak and Zhitnitsky \cite{Chernyak:1983ej}, QCD Sum Rules by Ball and Zwicky \cite{Ball:2004ye}, light-front quark model of Choi and Ji with two different potentials \cite{Choi:2007yu}, renormalon model of Agaev \cite{Agaev:2005rc}, instanton vacuum models of Petrov et al. \cite{Petrov:1998kg} and Nam et al. \cite{Nam:2006au}, Dyson-Schwinger Equations of Chang et al. \cite{Chang:2013pq} in the rainbow-ladder (RL) approximation and using the dynamical chiral symmetry breaking improved kernel (DB) and finally the platykurtic DA of Stefanis et al. \cite{Stefanis:2014nla,Stefanis:2014yha}.}
  \label{tab:DAmoments}
\end{table}
%%%%%%%%%%%%%%%%%%%%%%%%%%%%%%%%

\begin{figure}[htbp]
\centering 
\includegraphics[width=15cm,height=15cm]{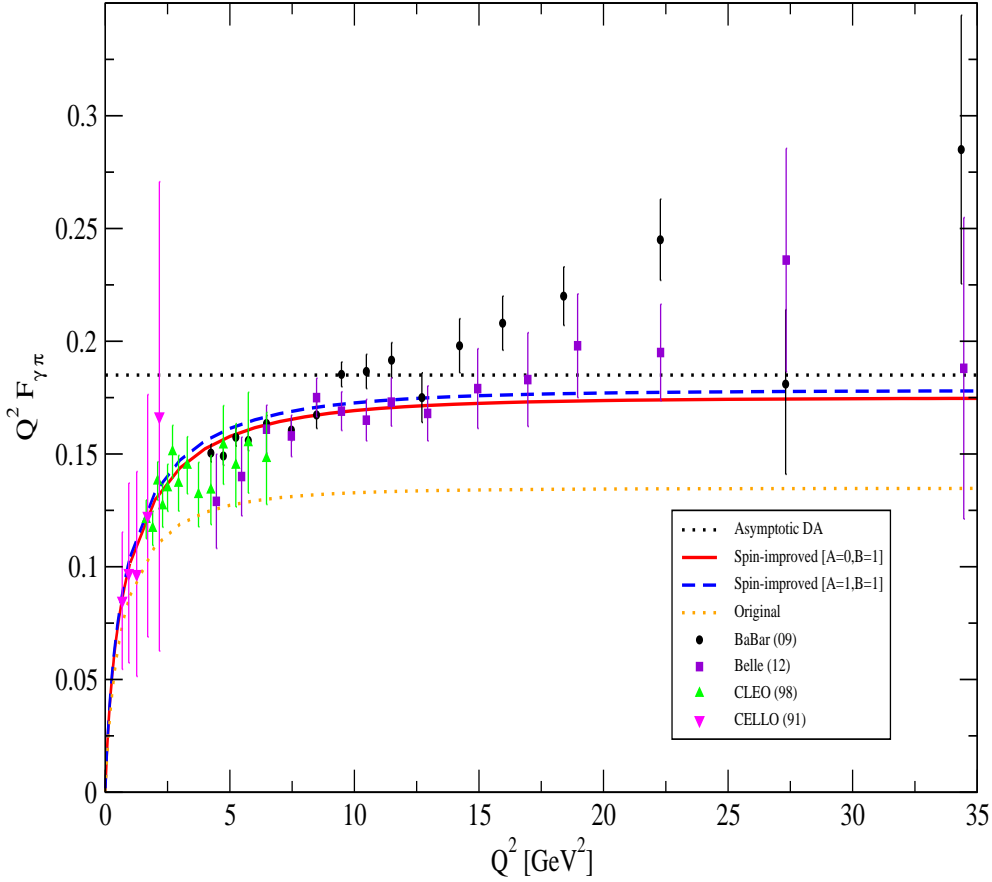}
\caption{Our predictions for the photon-to-pion TFF as a function of the photon's virtuality, $Q^2$. Dotted-orange curve:  original DA. Continuous-red and dashed-blue curves: spin-improved with $[A=0,B=1]$ and $[A=1,B=1]$ respectively. All theory curves are generated with $\kappa=523$ MeV and $m_f=330$ MeV. The dotted black curve is generated using the asymptotic DA with the measured pion decay constant. For asymptotic $Q^2$, they indicate the Brodsky-Lepage limit.  The data are from CELLO \cite{Behrend:1990sr}, CLEO \cite{Gronberg:1997fj}, BaBar \cite{Uehara:2012ag} and Belle \cite{Uehara:2012ag}.} 
\label{fig:TFF}
\end{figure}

Using our holographic DAs, we are able to predict the photon-to-pion transition form factor (TFF) which, to leading order in pQCD, is given as \cite{Lepage:1980fj}
\begin{equation}
	F_{\gamma \pi} (Q^2)= \frac{\sqrt{2}}{3} f_{\pi} \int_0^1 \mathrm{d} x \frac{\varphi_{\pi}(x,xQ)}{Q^2 x} \;.
	\label{TFF}
\end{equation}
We note that, even when computing the TFF in the perturbative region $Q^2 \ge 1~\text{GeV}^2$, the DA itself is probed at a scale $\mu=xQ$, which can be low if $x$ is close to its end-points.  Figure \ref{fig:TFF} shows that our spin-improved holographic DAs (continuous-red and dashed-blue curves) do a better job than the original holographic DA (dotted-orange curve). We note that the BaBar (2009) data \cite{Aubert:2009mc} indicate a strong scaling violation in disagreement with the Brodsky-Lepage limit: $F_{\pi \gamma^* \gamma}(Q^2 \to \infty)=\sqrt{2} f_{\pi}$ obtained by substituting the asymptotic DA in Eq. \eqref{TFF} and shown as the dotted-black curves in Figure \ref{fig:TFF}. The more recent Belle (2012) data \cite{Uehara:2012ag} do not confirm the BaBar (2009) data for $Q^2 > 10~\text{GeV}^2$ and the issue is likely to be resolved by precise future measurements in this kinematic range. Our spin-improved holographic wavefunctions clearly cannot describe the strong scaling violation indicated by the BaBar(2009) data and neither do our predictions exceed the asymptotic Brodsky-Lepage limit. For alternative models of the pion DA which are able to accommodate the BaBar (2009) data, we refer to Refs. \cite{Radyushkin:2009zg,Polyakov:2009je,Li:2009pr} and for an exhaustive analysis of the TFF data, we refer to \cite{Bakulev:2012nh}.

\section{Conclusions}
We have accounted for dynamical spin effects in the holographic pion light-front wavefunction and found a remarkable improvement in the description of pion radius, decay constant, EM form factor and photon-to-pion TFF. To generate our predictions, we have used a  constituent quark mass of $330$ MeV and the universal AdS/QCD scale $\kappa=523$ MeV, together with the assumption that the pion consists only of the leading quark-antiquark Fock state. Our results suggest that it could be possible to have a unified treatment of all light mesons, including the pion, with a universal fundamental AdS/QCD scale which fits the baryon and meson Regge slopes and also accurately predicts the non-perturbative QCD scale $\Lambda_{\text{QCD}}^{\overline{MS}}$. We also found that the predicted moments for the spin-improved holographic pion twist-$2$ DA are in better agreement with the predictions of standard non-perturbative methods such as lattice QCD and QCD Sum Rules, although they remain smaller than the latter. This suggests that tuning the values of $A$ and $B$ could be necessary or, at a deeper level, that the assumption underlying our Eq. \eqref{Spin-space} might not capturing all the dynamical spin effects in the pion. But this assumption, together with $(A=0,1;B=1)$, does bring a significant improvement in the description of all available experimental data without necessarily having to use a much smaller AdS/QCD scale and/or invoke higher Fock states contributions exclusively for the pion. Our findings thus support the idea of the emergence of a universal AdS/QCD confinement scale $\kappa$.

\section{Acknowledgements}
The work of M.A and R.S is supported by a team grant from National Science and Engineering Research Council of Canada (NSERC). F.C and R.S thank Mount Allison University for hospitality where parts of this work were carried out. We thank J. R. Forshaw, N. Stefanis and G. de T\'eramond for useful comments on the first version of this paper. 
\bibliographystyle{apsrev}
\bibliography{PionRevised3.bib}
\end{document}